\documentclass[final,3p,times,twocolumn]{elsarticle}

\usepackage{graphics}

\usepackage{amssymb}

\journal{Chemical Physics Letters}

\begin{document}

\begin{frontmatter}

\title{Spring-block approach for nanobristle patterns}

\author[ubb]{Ferenc J\'arai-Szab\'o}
\author[iwr]{Em\H{o}ke-\'Agnes Horv\'at}
\author[rpi]{Robert Vajtai}
\author[ubb]{Zolt\'an N\'eda}

\address[ubb]{Faculty of Physics, Babes-Bolyai University, 
              RO-400084 Cluj-Napoca, str. Kogalniceanu nr. 1, Romania}

\address[iwr]{Interdisciplinary Center for Scientific Computing, University of Heidelberg, 
              Speyererstr 6, 69115 Heidelberg, Germany}

\address[rpi]{Renseelaer Polytechnic Institute, Department of Materials Science and Engineering,
              Troy, NY, USA}

\begin{abstract}
A two dimensional spring-block type model is used to model capillarity driven self-organization 
of nanobristles. The model reveals the role of capillarity and van der Waals forces in the pattern formation
mechanism. By taking into account the relevant interactions several type of experimentally 
observed patterns are qualitatively well reproduced. The model offers the possibility to generate
on computer novel nanobristle based structures, offering hints for designing further experiments.  
\end{abstract}

\begin{keyword}
self-assembled nanostructures \sep spring-block models \sep pattern formation \sep nanotubes


\end{keyword}

\end{frontmatter}




Reproducible nanoscale patterns and structures are of wide interest nowadays for 
engineering components in modern small-scale electronic, optical and magnetic 
devices \cite{Rieth2003}. The so-called \emph{bottom up} approach for the fabrication
of these nanostructures uses nanoparticles as elementary building blocks. 
Under some specific conditions the nanoparticles self-organize into the 
desired structures \cite{Adachi2006}. A well-known and widely explored possibility
to induce this self-organization is to use the capillarity forces which appear 
during the drying of a liquid suspension of nanoparticles \cite{Haynes2001,Chabanov2004}. 
For instance, regular and irregular two-dimensional polystyrene nanosphere arrays on silica substrates
are generated by such methods \cite{Jarai2007}. These patterns are used then as a 
convenient mask in the NanoSphere Litography (NSL) method.

Carbon nanotubes (CNT) are attractive materials for nanotechnology because of their 
interesting physico-chemical properties and molecular symmetries. In order to make them 
appropriate for certain applications, proper initial CNT configurations have to be 
built, and specific conditions have to be found which enable their controlled 
self-organization \cite{Ding2007,Young2008,Meshot2009}. This is a very 
ambitious and challenging task, which can be 
made easier by elaborating working computer models for the self-organization of CNTs on substrates. 
Therefore, not only experimental, but also  
computational studies can advance the field of nanoengineering.

In the work of Chakrapani et al. \cite{Chakrapani2004} an experimental 
procedure is presented in which capillary self-organization of CNTs leads to puzzling cellular
patterns. After the collapse and reassembly of the highly ordered, anisotropic, 
elastic CNTs, shrinkage and crack formation occurs. The resulting 
cellular foams are visually striking, stable patterns.

\begin{figure}[tb!]
\centering%
\includegraphics[width=75mm]{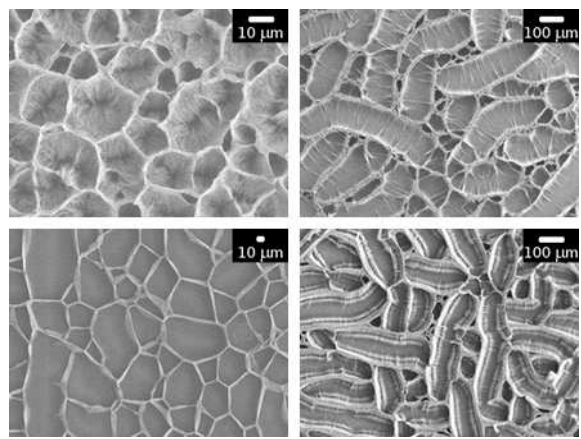}
\caption{Scanning electron microscope images of the structures obtained in 
         drying nanobristles.\label{fig1}}
\end{figure}

The experimental procedure \cite{Chakrapani2004} may be shortly summarized as 
follows. Multi-walled nanotube arrays are grown on rigid silica surface by chemical 
vapor deposition (CVD) based on the decomposition of ferrocene and xylene. The resulting 
nanotubes have a wall thickness of ca. 10nm and a diameter of ca. 30nm. The average 
distance between two nanotubes is ca. 50nm. The obtained nanotube bristle is 
oxidized in an oxygen plasma at room temperature and 133 Pa pressure for 10 minutes 
and immersed in a wetting fluid. After the liquid evaporates, characteristic cellular 
type patterns are formed, i.e. the ends of nanotubes self-organize in compact walls. 

Figure \ref{fig1}. shows scanning electron microscope images of some typical 
structures. From the figures we learn that a wide variety of structures are 
engineered in such manner. Both statistically symmetric polygonal cells and 
rather elongated ones can be obtained by chaning the experimental conditions. 

These micrometer scale structures have many advantageous features. They can be elastically deformed, 
transferred to other substrates or floated out to produce free-standing macroscopic 
fabrics. Thus, they might find potential applications as shock absorbent 
reinforcement in nanofiltration devices, elastic membranes and fabrics, and 
containers for storage or growth of biological cells. 

Despite of its applications and the existence of well elaborated production protocols,
the exact mechanisms responsible for self-organization of CNTs into vertically aligned 
cellular structures is not clearly understood. Recently, it has been argued that 
although we lack some basic information regarding the self-organization of CNTs 
within a bristle, this process can be approximated with the self-organization of arrays 
of CNT micropillars of micron-scale diameters \cite{DeVolder2010} each consisting of 
thousands of CNTs. This observation enables the construction of a computationally tractable 
model which operates instead of stand-alone CNTs with micropillars. 

In the present work a simple mechanical spring-block type model 
defined at mesoscopic micropillar level is considered for understanding the capillarity driven 
self-organization of nanobristles (or so called ``CNT forests''). The aim is to 
build a model that is able to qualitatively  reproduce the variety of experimentally produced patterns. 


The model used here for describing the self-organization process of nanobristles 
is a mesoscopic one and is based on the mechanical spring-block stick-slip model family. 
This model family appeared in 1967, when R. Burridge and L. Knopoff \cite{Burridge1967} constructed 
a simple mechanical model for explaining the Guttenberg-Richter law for the distribution of earthquakes 
after their magnitude. The basic elements of the model are blocks and springs that interconnect in a lattice-like 
topology. The blocks can slide with friction on a planar surface. The original system 
introduced by Burridge and Knopoff (BK) is a one-dimensional model. It can be 
studied numerically and it exhibits self-organized criticality \cite{Bak1996}.

The BK model gained new perspectives with the strong development of computers 
and computer simulation methods. Variants of the BK model proved to be useful in describing 
complex phenomena where avalanche-like processes are present, pattern formation phenomena 
and mesoscopic processes in solid-state physics or material sciences 
\cite{Andersen1994,Kovacs2005}.

Recently, by using this model, we have succesfully explained the patterns obtained in 
capillary self-organization of nanospheres \cite{Jarai2007,Jarai2005}. Motivated by this success, hereby we propose 
to map the capillarity driven self-organization of nanotube bristles to a spring-block system, 
and to understand the pattern selection process by means of computer simulations.

\begin{figure}[tb!]
\centering%
\includegraphics[width=75mm]{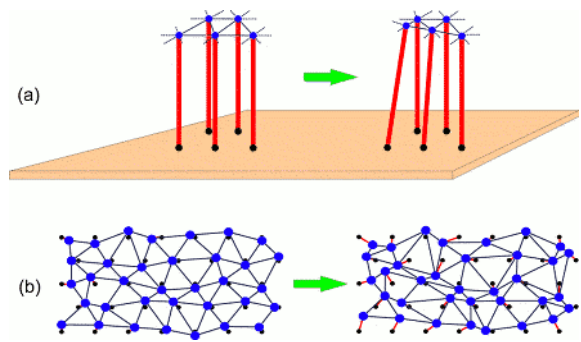}
\caption{Main elements of the spring-block model. Panel (a) shows a schematic 3D representation
of the nanobristle for the initial and a later state. Panel (b) illustrates the dynamics
of the equivalent 2D model. \label{fig2}}
\end{figure}

First, let us consider the three-dimensional (3D) model, which is very similar to the 
real nanotube arrangement. As sketched in Figure \ref{fig2}(a), the micropillars composed 
by thousands of nanotubes having fixed bottom ends are modeled by flexible strands. 
Their capillary interactions are represented by non-classical springs that connect
the neighboring pillars. The evaporation of the liquid is simulated by the stepwise 
increase of tension in the springs. This will result in the agglomeration of micropillar 
ends creating the final structure in the studied system.

As shown in Figure \ref{fig2}(b), this 3D model can be easily mapped 
into a two-dimensional (2D) one by projecting the micropillars' top ends on the surface.
In the projection plane the micropillars bottom ends are represented as
dots, and their positions are fixed on a predefined lattice. The movable top ends 
are modeled by the disk shaped blocks which can slide with friction on the 2D simulation surface. 
For visual purposes only, each block is 
connected by an extensible string with its bottom end showing the micropillars' trunk. In our simplest approach 
there is no restriction imposed to the length of these extensible 
strings which means that nanotubes with infinite length are used. This corresponds to the 
real case when the nanotubes length is much grater than the linear size of the cells in the final 
patterns. The blocks (top of micropillars) are connected with their nearest neighbors through 
special springs that model the resulting forces acting between the micropillars. 

\begin{figure}[tb!]
\centering%
\includegraphics[width=70mm]{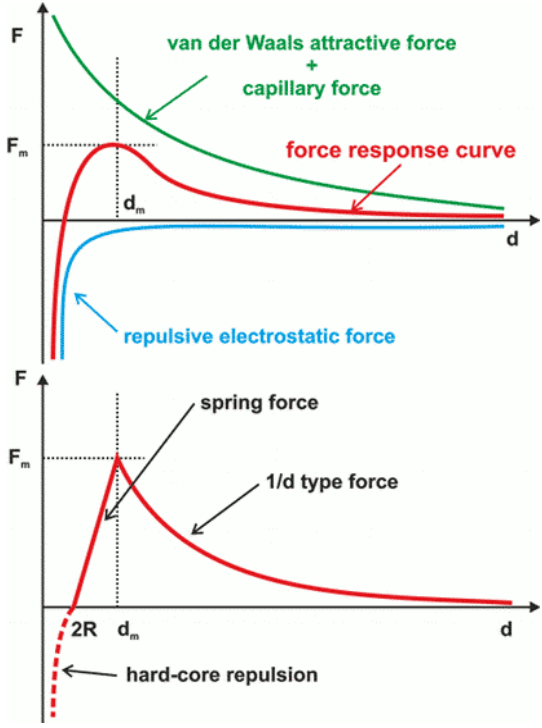}
\caption{Forces acting between micropillars immersed into liquid (top panel), 
the length dependence of the special spring force used for modeling this 
interaction (solid line on bottom panel), and the hard--core repulsion
force between blocks (dashed line on bottom panel)}.\label{fig3}
\end{figure}

These special springs are one key ingredient of our computational model. They
represent the resultant interaction force acting between two micropillars immersed in 
suspension. The tension force in the spring has a complex variation with the spring-length, it's form is sketched 
with a red line in the top panel of Figure \ref{fig3}. This force is
the resultant of the capillary force, the electrostatic repulsion and 
the van der Waals attraction between the micropillars. The form of the capillary force acting
between two micro-scale rods wetted by a liquid was measured experimentally and deduced analytically, 
too \cite{Dushkin1995,Kralchevsky1993,Kaptay2005}. Electrostatic repulsion and
van der Waals attraction are successfully described by the Derjaguin-Landau-Verwey-Overbeek
theory \cite{Hiemenz1997}. The capillary force decays inversely proportionally with distance
$F_c \propto 1/d$. The resultant of all these forces reduces to zero for long 
distances and it has a maximum $F_m$ at distance $d_m$ of micropillars.

Our model approximates this resulting force. The special spring-force used in 
simulations is shown with solid line in the bottom panel of 
Figure \ref{fig3}. For small elongations
$d < d_m$ this spring acts as a classical spring with
\begin{equation}
F_k (d) = k (d - 2R)\,,
\end{equation} 
where $k$ is the spring constant and $2R$ is its equilibrium length. 
For elongations longer than $d_m$ the spring force decays as 
\begin{equation}
F_k(d) = k' / d\,,
\end{equation}
where the 
constant $k'$ is selected in such way that the force-elongation curve is continuous at $d_m$. 
The maximum of the spring force at $d_m$ is denoted by $F_m$.

Similarly with our previous models of drying granular materials \cite{Leung2000} or 
self-organizing nanosphere systems \cite{Jarai2005}, the effects created by the evaporation of the liquid 
is introduced through these 
springs. As the liquid evaporates the meniscus accounting for the capillarity forces 
gets more accentuated. This is modeled by a step-by-step increasing of the spring
constant $k$. It has to be noted that through the drying process the maximum force 
in the spring $F_m$ remains unchanged since the electrostatic forces are not affected. 
Accordingly, with the increasing of $k$ the $d_m$ value has to be proportionally lowered.

The second key ingredient of our spring-block model (necessary to get realistic structures) 
is the capillary force resulting from the non-vertical orientation of 
micropillars. Once the micropillar becomes inclined, the meniscus radius of the liquid 
surface in contact with the micropillar becomes grater on the top side than on the bottom side.  
This leads to an unstable state because a resulting net force $F \sim 1/cos \theta$ pointing vertically 
downwards will act at the crossing point between the liquid surface and the micropillar 
\cite{Neukirch2007}, tending to incline even more the tube. Here we denoted by $\theta$ the angle between 
the vertical and the tangent to the micropillar at the liquid level. 

In our spring-block approach this force which is monotonically increasing with the inclination angle 
is approximated by a simple linear repulsion force 
\begin{equation}
F_a(x) = k_a x
\end{equation} 
acting between the 
block and its fixed bottom end ($x$ denotes their distance in the simulation 
plane and $k_a$ is the repulsion constant).

There is an additional almost hard-core--type repulsion $F_j$ which forbids blocks to 
interpenetrate. This is taken into account by the repulsive part of a Lennard--Jones type 
force which acts only when the distance between two blocks becomes smaller than $2R$
(dashed line on the bottom panel of Figure \ref{fig3}).

Additionally to the presented forces, damping forces are considered to stabilize the dynamics. 
A friction (pinning) between blocks and surface is introduced. It can equilibrate a net force less than $F_p$. 
Whenever the total force acting on a block exceeds $F_p$, the block slips with an over-damped
motion.

The dynamics leading to pattern formation consists of the following relaxation steps. 
First, the system is initialized. Blocks are placed on a triangular 
lattice with lattice constant $a$ and their corresponding bottom ends are fixed at the 
same positions. Then, the blocks are slightly dislodged in a random direction with a 
small random shift not grater than the half of the empty space between blocks. Thereby, 
the initial imperfectness of the nanobristle is modeled. Next, the interconnecting 
spring-network is build. The spring constant $k$ is selected in such way that in 
the initial system the spring forces are not exceeding the pinning force $F_p$ that acts 
on each block. An initially prestressed spring-block network is thus constructed.

During each simulation step the spring constant is increased by a small amount $\delta k$ 
representing the increase of tension due to the evaporation of water and the system relaxes 
to an equilibrium configuration. In this configuration the total net force acting on each disk is lower in 
magnitude than the pinning threshold $F_p$. 

Similarly with the modelling of drying nanosphere suspensions \cite{Jarai2005}, the
relaxation dynamics is realized through an over-damped molecular dynamics simulation using a
fixed time-step. The equilibrium state in a viscous medium can be found by moving the 
blocks in each simulation step in the direction of the net force acting on them, and with a 
displacement which is proportional to the magnitude of the resultant force
\begin{equation}
d \mathbf r=\frac {1}{\nu} \mathbf F(\mathbf r) dt,
\label{overdamped}
\end{equation}
where $\nu$ denotes a viscosity. 

Since the simulation steps will be chosen small enough, this simplified dynamics is 
able to replace the Newtonian solution without loss of significant information. We remind 
here that we are not interested in the dynamics, but in the final equilibrium configuration.
A relaxation step is finished when no disk slipping event occurs for the given spring 
constant value. It usually takes a very long time to achieve a perfect relaxation, therefore we introduce
a tolerance ($10^{-6}$--$10^{-9}$), and assume that the relaxation is completed when the largest
displacement per unit time is smaller than this value. After relaxation is done, we proceed 
to the next simulation step and increase all spring constants by $\delta k$. This dynamics is 
repeated until a final, stable configuration of the blocks is reached.

When implementing the above relaxation dynamics, several types of boundary conditions can be considered. 
However, as it was shown in \cite{Jarai2005} the boundary conditions (free, fixed or periodic) 
will influence the final stable structure only in the vicinity of boundaries. In the bulk, the 
obtained structures are rather independent of this choice. Accordingly, in the present simulations 
fixed boundary conditions are used and snapshots from the bulk are taken for later investigations. 

The above presented model has many parameters. Here, their values for our 
present simulations are given. The disk shaped blocks are considered with radii $R = 1$, and it 
defines the unit length in the system. After fixing the diameter of the circular simulation area 
$D = 400 - 600$, the blocks 
are placed on the triangular lattice having a lattice constant $a = 2.2 - 5.0$. The density 
(or space filling) of micropillars is implicitly defined by this lattice constant. The block 
sliding dynamics is governed by the viscosity $\nu = 250$ used in the equation (\ref{overdamped}) 
and the pinning threshold $F_p = 0.001$.
The springs used in simulations are characterized by two parameters, namely the
initial spring constant $k = 0.01 - 0.05$ and the equilibrium distance of springs $2R$.
By these parameters the unit force in our model system is defined. In order to simulate a 
quasi-static drying process the spring constant increasing step has to be a small one 
$\delta k = 0.001$. The central capillary repulsion constant $k_a = 0.0003$ is set to be 
small enough that in the initial system, where the micropillars are almost vertical, 
not to affect the cell formation dynamics. Later, when the walls are forming (and the 
micropillars are no longer vertically aligned), this force becomes grater and it helps 
the wall formation and stabilization process. The used Lennard--Jones force has its standard 
parameters set to $\sigma = 1.79$ and $\varepsilon = 1.3 10^{-7}$ expressed in simulation units.
All results presented in the following are obtained with the above choosed parameter values, 
unless it is otherwise specified in figure captions. 


\begin{figure}[tb!]
\centering%
\includegraphics[width=75mm]{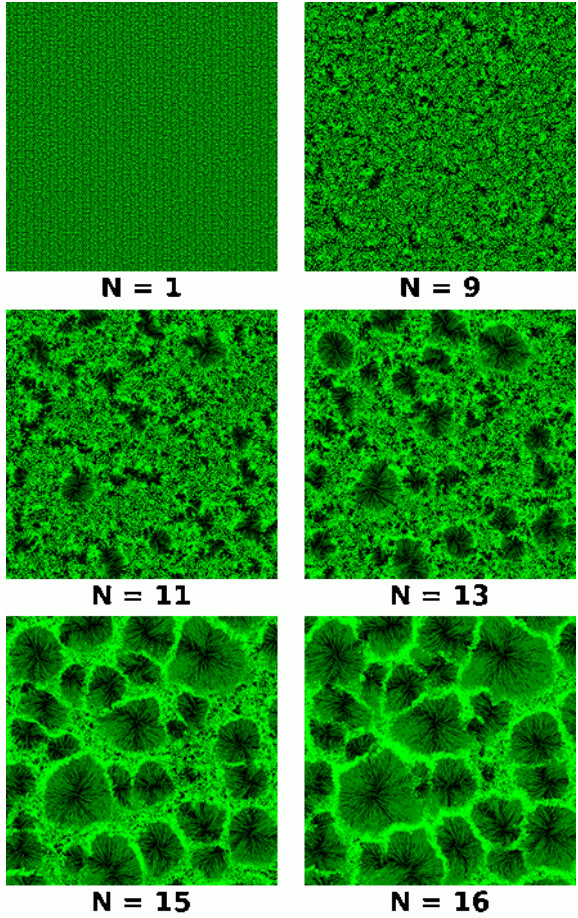}
\caption{Time evolution of the simulation for parameters $D = 400$, $a = 2.2$ and $k = 0.01$.
The simulation time step $N$ is noted below the snapshots.\label{fig4}}
\end{figure}

The presented algorithm can be easily implemented and systems up to 300\,000 nanotubes can
be simulated in reasonable computational time. As observable from the time-sequence in Figure \ref{fig4}, 
the cellular patterns are formed after nucleation of voids in the spring-block network 
(time step $N = 9$). A preliminary void is enlarged by the tensioned 
springs (time step $N = 13$) until the top of micropillars arrange in a final and 
stable cellular structure (time step $N = 16$).
The obtained dynamics resembles the pattern formation mechanism known from experimental
in-situ observations \cite{Chakrapani2004}.

\begin{figure}[tb!]
\centering
\includegraphics[width=75mm]{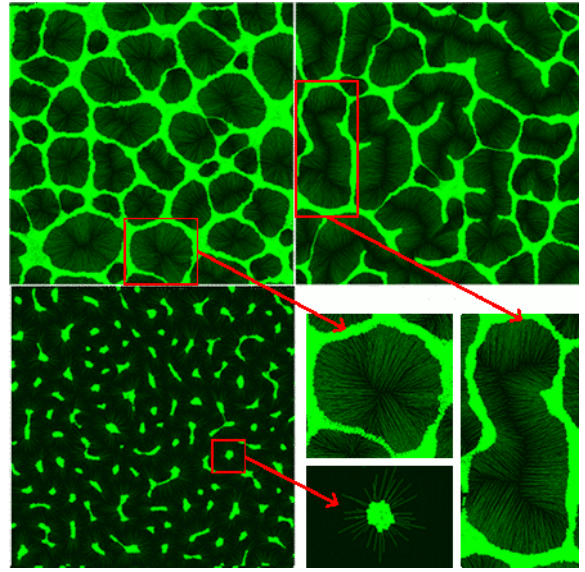}
\caption{Different final structures depending on micropillar density obtained by 
         simulations with parameters $D = 600$ and $k = 0.05$. The lattice constant is $a = 2.4$ for the
         top left panel, $a = 3.0$ for the top right panel and $a = 5.0$ for the bottom left panel. 
         The obtained structures are magnified on the bottom right panel.\label{fig5}}
\end{figure}

Furthermore, the effect of nanotube density on final patterns is computationally investigated.
Simulations with the same parameter set are preformed for systems initialized with different lattice 
constants $a$. By this, the space filling of the micropillar system 
\begin{equation}
\rho = \frac{2\pi}{a^2 \sqrt{3}}
\end{equation}  
is varied. 
Here, it has to be noted that the micropillar density and implicitly the 
micropillar lattice constant are linearly connected to  the real nanotube density $\rho_n$ and 
the nanotube lattice constant $a_n$, respectively. If a micropillar is composed by $N$ nanotubes, 
then by simple geometrical calculations it can be shown that 
\begin{equation}
	\rho = \rho_n \frac{\pi R^2}{2 R_n^2\sqrt{3}N}\,\, {\rm and}\,\, 
	a = a_n \sqrt{\frac{2\sqrt{3}N}{\pi}}\,,
\end{equation}
where $R_n$ denotes the radius of a nanotube.

In Figure \ref{fig5} three different type of simulated structures are presented. For high space filling 
$\rho = 0.688$ corresponding to lattice constant $a = 2.4$ polygonal cellular structures are obtained 
similar to those on the left hand side of Figure \ref{fig1}. As one can observe on the magnified cell 
image, at the center there is a clean area formed by radially outgoing micropillars. For intermediate 
space filling $\rho = 0.404$ which corresponds to a lattice constant $a = 3.0$ the micropillars 
self-organize into elongated cellular structures. The obtained structures are in qualitative agreement 
with the experimental structures presented in the right hand side SEM images of Figure \ref{fig1}.
When interpreting the image, one has to take
into account that the length scale of this snapshot is $1.25$ times smaller than the length scale of 
the previously discussed one. Accordingly, the typical size of the simulated elongated structures is 
$3-5$ times grater than the typical size of the polygonal structures. For low space filling 
$\rho = 0.145$ corresponding to a lattice constant $a = 5.0$ the micropillars form bundled clusters
resembling the recent experimental patterns of sequential assembly of self-similar nanotube clusters 
\cite{Pokroy2009}.

\begin{figure}[tb!]
\centering%
\includegraphics[width=75mm]{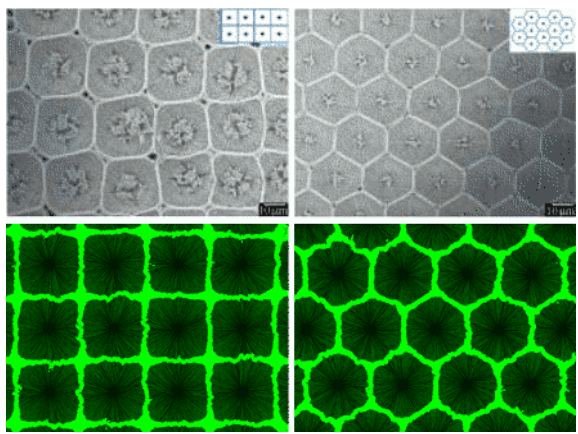}
\caption{Highly ordered experimentally designed patterns \cite{Liu2004} (top) in comparison with 
         simulation results (bottom) for the parameter set $D = 600$, $a = 2.6$ and $k = 0.01$. \label{fig6}}
\end{figure}

Finally our simulations explored also the possibility of designing highly ordered micropatterns 
in nanobristles. In this sense, an experimental procedure elaborated in the last years \cite{Liu2004}
has been reconstructed by our spring-block type simulations. The design procedure is based on 
the experimental observation that low-density regions or vacancies in the bristle play an important 
role in the pattern cell nucleation process.

As shown in top panels of Figure \ref{fig6}, different kinds of highly ordered micropatterned 
structures have been created by etching regular vacancies with laser pulses. Capillary self-organization 
in such systems yields the structures presented on the top panels of Figure \ref{fig6}. 

Simulations on micropillar forests with vacancies of radius $r = 5$ created on rectangular and triangular
lattice have been performed. The results are shown on bottom panels of Figure \ref{fig6}. 
In agreement with experimental findings, our spring-block simulation results suggest that various 
micropatterns may be designed by proper preparation of the initial nanobristle taking into account
that the wall of a polygon shaped cell forms approximately at the vertical bisector of two adjacent 
vacancies.


In conclusion, a simple mechanical, spring-block type model has been proposed here to model
capillarity driven self-organization of nanobristles. The 3D problem has been mapped to a 
2D model that works at the mesoscopic micropillar level incorporating real interactions 
known from earlier experimental findings. Our computer simulations evidenced the role of capillary 
and electrostatic forces in forming the self-organized nanostructures.  The dynamics leading 
to pattern formation has been also revealed. By using the same model parameters with different 
nanotube densities three qualitatively different types of patterns were reproduced. Moreover, 
the possibility of designing highly ordered nanostructures has also been explored.

\section*{Acknowledgement}
This work was supported by CNCSIS-UEFISCSU, project number PN II-IDEI 2369/2008. 
One of the authors
(E-\'AH)  was partly funded by a scholarship from  the  Heidelberg Graduate School of Mathematical 
and Computational Methods for the Sciences, University of Heidelberg, Germany, which is funded by 
the German Excellence Initiative (GSC 220).

\bibliographystyle{elsarticle-num}

\end{document}